\documentclass{aa}
\usepackage{graphicx}
\usepackage{txfonts}
\begin{document}
   \title{First results from the INTEGRAL Galactic plane scans\thanks
   {Based on observations with INTEGRAL, an ESA project with instruments 
   and science data centre funded by ESA member states (especially the PI 
   countries: Denmark, France, Germany, Italy, Switzerland, Spain), Czech 
   Republic and Poland, and with the participation of Russia and the USA.}
   }


   \author{C. Winkler\inst{1}, N. Gehrels\inst{2},  
   V. Sch\"onfelder\inst{4}, J.-P. Roques\inst{6},
  A.W. Strong\inst{4}, C. Wunderer\inst{4}, P. Ubertini\inst{5}, \\F. Lebrun\inst{9}, A. Bazzano\inst{5}, M. Del Santo\inst{5}, N. Lund\inst{3},
   N.J. Westergaard\inst{3}, V. Beckmann\inst{7}$^,$\inst{10}, 
P. Kretschmar\inst{7}$^, $\inst{4}\\
          \and
	  S. Mereghetti\inst{8}
          }

   \offprints{C. Winkler}

   \institute{ESA-ESTEC, Research and Scientific Support Department, Keplerlaan
   1, NL-2201 AZ Noordwijk, The Netherlands, 
              \email{Christoph.Winkler@rssd.esa.int}
	      \and
	      NASA-GSFC, Code 661, Greenbelt, MD 20771, USA,
	      \email{gehrels@gsfc.nasa.gov}
	      \and
	      DSRI, Juliane Maries Vej 30, DK-2100 Copenhagen OE,
	      Denmark,
	      \email{nl@dsri.dk, njw@dsri.dk}
	      \and
	      MPE, Postfach 1603, D-85740 Garching, Germany,
	      \email{vos@mpe.mpg.de, aws@mpe.mpg.de, cow@mpe.mpg.de}
              \and
             IASF-CNR, Via del Fosso del Cavaliere 100, I-00133 Rome, Italy,
             \email{ubertini@rm.iasf.cnr.it}
	     \and
             CESR, Boite Postale 4346, F-31029 Toulouse, France,
	     \email{roques@sigma-0.cesr.cnes.fr}
	     \and
	     ISDC, Chemin d'Ecogia 16, CH-1290 Versoix, Switzerland,
	     \email{Volker.Beckmann@obs.unige.ch\\
Peter.Kretschmar@obs.unige.ch}            
             \and
             CNR-IASF, via Bassini 15, I-20133 Milano, Italy,
             \email{sandro@mi.iasf.cnr.it}
             \and
              Service d'Astrophysique, CEA-Saclay, F-91191 Gif-sur-Yvette Cedex, France,
             \email{flebrun@cea.fr}
             \and
             Institut f\"ur Astronomie und Astrophysik, Universit\"at T\"ubingen, Sand 1, D-72076 T\"ubingen, Germany
}

   \date{Received ; accepted}

   \abstract{
   Scans of the Galactic plane performed 
   at regular intervals constitute a key element of the guaranteed time 
   observations of the INTEGRAL 
   observing programme. These scans are done for two reasons: frequent
   monitoring of the Galactic plane in order to detect transient sources, and
   time resolved mapping of the Galactic plane in continuum and diffuse line
   emission. This paper describes first results obtained from the Galactic
   plane scans executed so far during the early phase (Dec 2002 - May 2003) of the nominal mission.
   
   }

   \authorrunning{C. Winkler et al.} 
   \titlerunning{INTEGRAL Galactic Plane Scans}

   \maketitle
%

\section{Introduction}
 
The Core Programme (CP) of guaranteed time observations (Winkler \cite{winkler}) during the first year of INTEGRAL's nominal mission consists of a deep exposure of the central Galactic radian (GCDE), regular scans of the Galactic plane (GPS) and pointed observations. A total observing time of 9.3 Ms -- corresponding to 35\% of the observing time available during the first year -- has been allocated to CP observations, out of which 2.3 Ms have been reserved for regular scans of the Galactic plane (GPS).

The GPS are mainly done for two reasons: 
the most important one is to provide frequent monitoring of the Galactic 
plane in 
order to detect transient sources, because the gamma-ray sky in the INTEGRAL 
energy range is dominated by the extreme variability of many sources
as demonstrated by previous X$-$ and gamma$-$ray missions. The 
scans would find sources in high state (outburst) which warrant possible 
scientifically important follow-up (Target of Opportunity) observations. 
The second reason is to build up 
time-resolved maps of the Galactic plane in 
continuum and diffuse line emission such as $^{26}$Al, $^{44}$Ti and 511 keV with modest 
exposure. 
CGRO and SIGMA detected Galactic compact sources of 
several different categories/groups which include 
X-ray binaries (e.g. X-ray novae, Be binary pulsars) and in particular
superluminal sources (GRS 1915$+$105 and GRO J1655$-$40). 
The pre-launch estimate on the occurrence rates for events that INTEGRAL can observe 
is about 2 events/year for 
each of these classes, where pointing constraints 
due to the fixed solar arrays of the INTEGRAL spacecraft (Jensen et al. \cite{jensen}) have been taken into account. The important 
time scales for the transient outbursts vary significantly from class to 
class and from event to event, but a typical duration of an event is 1 -- 2 
weeks and a typical variability time scale is of the order of 1 day. 
   
\begin{table*}[htp]
 
      \caption[]{Journal of Galactic plane scans, executed during early mission phase.}
         \label{GPS}
    
        \begin{tabular}{rcccrr}
            \hline
        \noalign{\smallskip}
 Exposure$^{\mathrm{a}}$ [seconds]& Start & End & \# Pointings &  Start (l, b) [deg]& Stop (l, b) [deg] \\
	              \noalign{\smallskip}
            \hline
            \noalign{\smallskip}
     24200    &      28-DEC-02 & 28-DEC-02 &      11  &   128.3, 2.1 &    71.9,  6.4 \\
     24200   &       30-DEC-02 & 30-DEC-02 &      11  &   73.3,   2.1   & 129.8, $-$2.1  \\ 
     33000   &       11-JAN-03 & 14-JAN-03 &      15  &   270.0, $-$2.1  & 123.4,  2.1   \\
     28600   &       29-JAN-03 & 29-JAN-03 &      13  &   337.0,  4.3   & 269.3,  4.3 \\
     59400   &       02-MAR-03 & 03-MAR-03 &      27  &   325.0, $-$2.1  & 189.6, $-$2.1 \\ 
    110000   &       12-MAR-03 & 16-MAR-03 &      50  &   54.6,  0.0   & 99.9,  4.3   \\
    123200   &       24-MAR-03 & 29-MAR-03 &      56  &   295.5,  4.3   & 110.4, $-$2.1 \\
     92400   &       05-APR-03 & 08-APR-03 &      42  &   273.3, $-$2.1  & 339.8, $-$2.1 \\
     83600   &       17-APR-03 & 20-APR-03 &      38  &   108.9,  2.1   & 278.1, $-$2.1 \\
     66000   &       28-APR-03 & 02-MAY-03 &      30  &   209.6,  4.3   & 102.5, $-$2.1 \\
     26400   &       13-MAY-03 & 14-MAY-03 &      12  &   226.0, 0.0 &  288.1, $-$2.1 \\
     28600   &       22-MAY-03 & 23-MAY-03 &      13  &   123.7,  2.1   & 56.0,  2.1 \\

%
%

            \noalign{\smallskip}
            \hline
         \end{tabular}

 \begin{list}{}{}
 \item[$^{\mathrm{a}}$] Note that the scheduled exposure is shown. Actual exposures obtained for IBIS and JEM-X during this period amounts to $\sim$95\% and to $\sim$75\% for SPI, due to telemetry constraints during the early mission phase.
 \end{list}

   \end{table*}
   \begin{figure*}[htp]
   \centering
   \includegraphics[width=\textwidth]{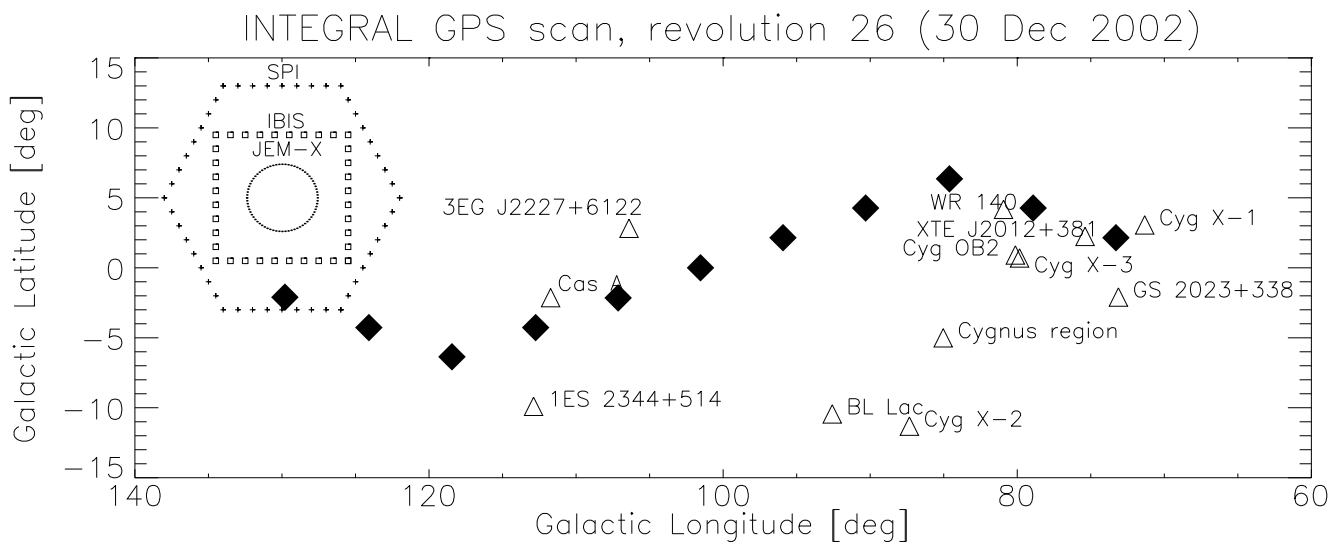}
      \caption{Pointings of the ''sawtooth-shaped'' GPS scan (filled symbols) as performed during revolution 26 (30 Dec. 2002). 
              The fully coded fields of view (FCFOV, SPI: 16$^{\circ}$, IBIS: 
9$^{\circ}$$\times$9$^{\circ}$, JEM-X: 5$^{\circ}$),  
              and some known high energy targets  
              are shown for illustration. 
              }
         \label{gpssample}
   \end{figure*}



The scans  occur
once per 4 revolutions (i.e. every 12 days) by performing a ``slew -- and stare'' manoeuvre 
of the spacecraft along the visible (accessible) part of the 
Galactic plane with latitude extent $\pm$10$^{\circ}$.
A ''revolution'' corresponds to the duration ($\sim$3 days) which INTEGRAL needs to complete one orbit.
The accessible part of the Galactic plane depends on viewing constraints 
including the 40$^{\circ}$ solar aspect angle due to the fixed solar arrays.
The angular distance 
between subsequent exposures (2200 s each) along the scan path is 6$^{\circ}$. 
The scans are performed as a ''sawtooth'' (Fig~\ref{GPS})
with inclination of 21$^{\circ}$ with respect to the Galactic plane, 
each subsequent scan is shifted by 27.5$^{\circ}$ in galactic longitude. The sawtooth patterns build up
a grid of various dither pointings, facilitating the imaging analysis for SPI.
The visibility of the Galactic plane depends on the annual season. On average
a strip of width b = $\pm$10$^{\circ}$ has a longitude 
extent of 170$^{\circ}$ (ranging from
110$^{\circ}$ to 360$^{\circ}$) and requires on average 21 hours of scanning
(ranging from 12 hours to 42 hours) which includes the total of exposures and spacecraft slews. In this paper we describe preliminary results obtained from GPS scans during the early mission phase from December 2002 to  May 2003 concentrating on the monitoring of known and the detection of new point sources.The GPS scans executed during the early phase of the mission  are shown in Table~\ref{GPS}.

\section{First results from SPI }

Two analysis tools have been developed for SPI (Vedrenne et al. \cite{vedrenne}) to analyse gamma$-$ray
sources: {\em Spiros} (Skinner et al. \cite{skinner}), which is specialized for point sources, and
{\em Spiskymax} (Strong et al. \cite{strong}) which is aimed more at diffuse emission but also images
point sources. {\em Spiros} uses an input catalogue and tests each source for significance
and flux.  It also seaches for sources in addition to those in the
catalogue.

The number of known sources found at significance level $>$5$\sigma$ by {\em Spiros} for
all scans during revolutions 25$-$67 (December 2002$-$May 2003) in the energy range (20$-$40) keV (using data from 234 pointings and 538 ks exposure) is 33, three yet unknown sources were found which need further study.
Well$-$known and bright sources are cleary detected in the short GPS
exposures; the status of the less significant sources has to be
investigated in detail on a source$-$by$-$source basis.  The same applies
to the 3 sources found which were not in the input catalogue. 
%
   \begin{figure*}
   \centering
   \includegraphics[width=\textwidth]{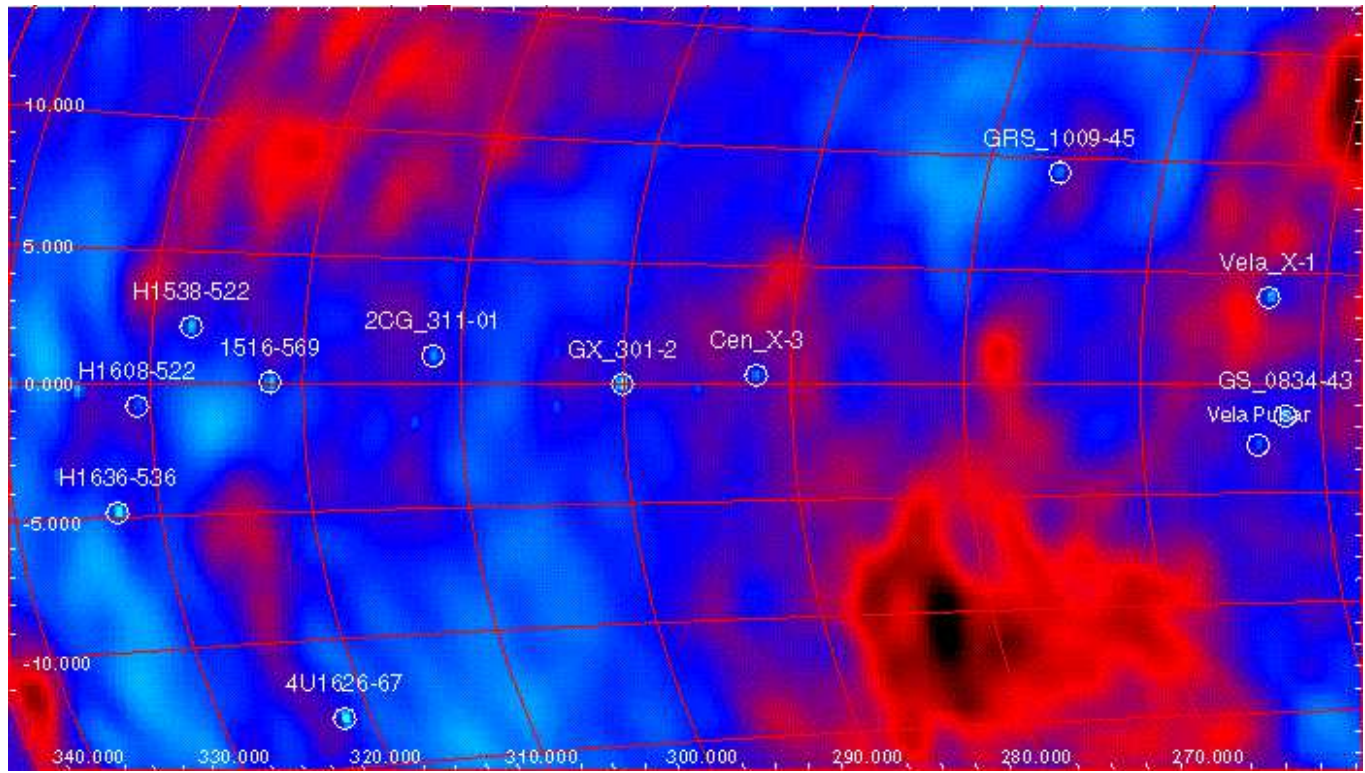}
      \caption{ Example of a SPI significance 
({\em Spiros}) sky map in galactic coordinates, (20$-$40 keV), obtained during GPS scans in the period December 2002 -- March 2003. 
              }
         \label{spimap}
   \end{figure*}
%
%

For SPI an important use of the GPS will be to extend the core programme's deep exposure (4.3 Ms) of the 
Galactic Centre (GCDE) to the
entire Galactic plane for large$-$scale mapping. Already with 6 months
data the plane coverage is almost complete, although at lower
exposure than the GCDE.  It enables the first sky maps of the whole
plane to be made.
Fig~\ref{spimap} shows a SPI significance skymap from the GPS derived using
{\em Spiros} in the (20$-$40) keV energy band . The bright sources detected during GPS are visible.
 
Another important scientific objective for SPI is to search for transient line features like
those discovered by SIGMA from Nova Muscae in 1991 (Goldwurm et al. \cite{goldwurm})  and from 1E1740.7$-$294 in 1990
(Bouchet et al. \cite{bouchet}, and 
Sunyaev et al. \cite{sunyaev}) and in 1992 (Cordier et al. \cite{cordier}).
Broad, transient
features (FWHM $\sim$ 50 -- 250 keV) centered  
between 350 keV and 500 keV were observed on timescales of $\sim$ hours 
to days.

   \begin{table*}[htp]
   \caption[]{SPI estimated sensitivities (during one GPS scan) to broadened line features at (350--500)~keV as observed by SIGMA.}
\label{gps_spi_annihilTransient}
        \begin{tabular}{llllll}
            \hline
Reference & Line  & Line  & SPI sensitivity      & SPI sensitivity      & Observed\\
     & Position  & FWHM  & (optimum)      & (intermed.)    & Flux    \\
     & [keV] & [keV] & [ph/(cm$^2$s)] & [ph/(cm$^2$s)] & [ph/(cm$^2$s)] \\

 \noalign{\smallskip}
            \hline
            \noalign{\smallskip}

\multicolumn{6}{c}{\em{1E1740.7 - Oct. 1990 event}} \\

Bouchet et al. \cite{bouchet}& 480   & 240  &7.7$\times10^{-3}$&10$\times10^{-3}$&13$\times10^{-3}$\\

Sunyaev et al. \cite{sunyaev}& 410   & 180
&6.2$\times10^{-3}$&8.4$\times10^{-3}$&(9.5$\pm4.5)\times10^{-3}$\\
\hline

\multicolumn{6}{c}{\em{1E1740.7 - Sep. 1992 event}} \\

Cordier et al. \cite{cordier} & 350   & 170
&5.5$\times10^{-3}$&7.5$\times10^{-3}$&4.3$^{+2.7}_{-1.5}\times10^{-3}$\\

\hline
\multicolumn{6}{c}{\em{1E1740.7 - Oct. 1990 $+$ Sep. 1992 events}} \\

Cordier et al. \cite{cordier}& 390   & 190  &6.5$\times10^{-3}$&8.8$\times10^{-3}$&\\

\hline
\multicolumn{6}{c}{\em{Nova Muscae - Jan. 1991 event}} \\

Goldwurm et al. \cite{goldwurm}& 481   & 54
&3.9$\times10^{-3}$&5.3$\times10^{-3}$&(6.0$\pm2.9)\times10^{-3}$\\
  
\noalign{\smallskip}
            \hline
         \end{tabular}
     
    \end{table*}

We have estimated the SPI sensitivity to such events during a single GPS
scan; depending on the source position, two or three consecutive GPS 
pointings (2200~s each) can be used. Table~\ref{gps_spi_annihilTransient}
lists calculated SPI sensitivities for the transient features based on quoted
line centroid and FWHM. Reported fluxes are shown for comparison.
We show two sets of estimated SPI sensitivities, one for optimum source location
in the path of the GPS scan, and one for an intermediate case. Of course, the
larger the distance of the source from the GPS scan path, the lower the sensitivity. The 
sensitivity estimates are based on single-detector events only (``SE+PE'') 
using background measured during revolutions 11 -- 45 (November 2002 -- Feb 2003). The mask opacity is 
approximated at 50\%, and we assume perfect background knowledge (i.e. background uncertainty
only from measurement statistics), and assume no other significant sources in the relevant energy band in the FOV.


\section{First results from IBIS }

IBIS (Ubertini et al. \cite{ubertini})
was designed as a gamma-ray wide field of view telescope with high sensitivity even for short observing time of the order of a few thousand seconds. This characteristic was optimised with the scientific goal to regularly monitor a large fraction of the Galactic plane and to discover most of the expected transient sources,  whose existence was anticipated by X--Ray mission like BeppoSAX and RXTE, operating at lower energies. During the first part of the CP observation a sensitivity of about 20 mCrab at 50 keV has been achieved for the GPS scan, combining together different pointings of the same sky region within the FOV. 

So far almost 100 known sources have been detected, including all persistent though variable LMXB and HMXB and several different flaring or out--bursting sources with intensity above 20--50 mCrab. As anticipated most of them are Neutron Star systems, either weakly or strongly magnetised and a few are  Black Hole candidates. At the same time  10 new INTEGRAL sources have been discovered (see Table~\ref{new}) by the IBIS low energy detector ISGRI. 
   \begin{table*}[htp]
      \caption[]{New INTEGRAL sources$^{\mathrm{a}}$.}
         \label{new}
     
        \begin{tabular}{llll}

            \hline
            \noalign{\smallskip}
           Source &  Flux (mCrab) & Flux (mCrab)  & Reference  \\
                  &     (15$-$40) keV   &  (40$-$100) keV                            &\\
           \noalign{\smallskip}
            \hline
            \noalign{\smallskip}

IGR J18483$-$0311 & 10   (S/N $\sim$21)  &   5 (S/N $\sim$11)&   2 May 2003 (ATEL 157) \\
IGR J17597$-$2201  & 5  (S/N $\sim$10)  &   10 (S/N $>$14)&  30 Apr 2003 (ATEL 155) \\
IGR J18325$-$0756 & 10             &     5      &      28 Apr 2003 (ATEL \#154) \\
IGR J18539$+$0727 & 20              &   20       &     21 Apr 2003 (ATEL \#151) \\
IGR J17091$-$3624 & --              &   20       &     19 Apr 2003 (ATEL \#149) \\
IGR J17464$-$3213 & 60              &   60       &     28 Mar 2003 (ATEL \#132) \\
IGR J16358$-$4726 & 50             &   20       &     19 Mar 2003 (IAUC \#8097) \\
IGR J19140$+$098  & 50$-$100          &            &     6 Mar 2003 (IAUC \#8088) \\
IGR J16320$-$4751 & 10$-$50          &            &      1 Feb 2003 (IAUC \#8076) \\
IGR J16318$-$4848 & 50$-$100        &            &     29 Jan 2003 (IAUC \#8063) \\
  
\noalign{\smallskip}
            \hline
         \end{tabular}
     
\begin{list}{}{}
 \item[$^{\mathrm{a}}$] All IGR sources were discovered during core programme observations (3 sources during GPS, 5 sources during GCDE), except IGR J16320$-$4751 and IGR J19140$+$098 which were detected during open time pointed observations.
 \end{list}

    \end{table*}

Most of them show substantial emission in the lower energy range (15--40) keV, and, even if showing a relatively soft spectrum (cf. IGR J16358--4726), few very hard ones have sustained emission up to (150--200) keV (cf. IGR J17464--3213). Others show an unusually high intrinsic absorption (cf. IGR J16318--4848) up to 2$\times10^{24}$ N$_H$  suggesting  the existence of a new class of Galactic objects. They appear to be enshrouded by a dense envelope and visible only above (5--10) keV (Revnivtsev et al. \cite{revnivtsev}), which puts them 
at the limit of CHANDRA and XMM-Newton capability, and amenable to deep investigations with IBIS.   

\begin{figure*}
   \centering
   \includegraphics[width=\textwidth]{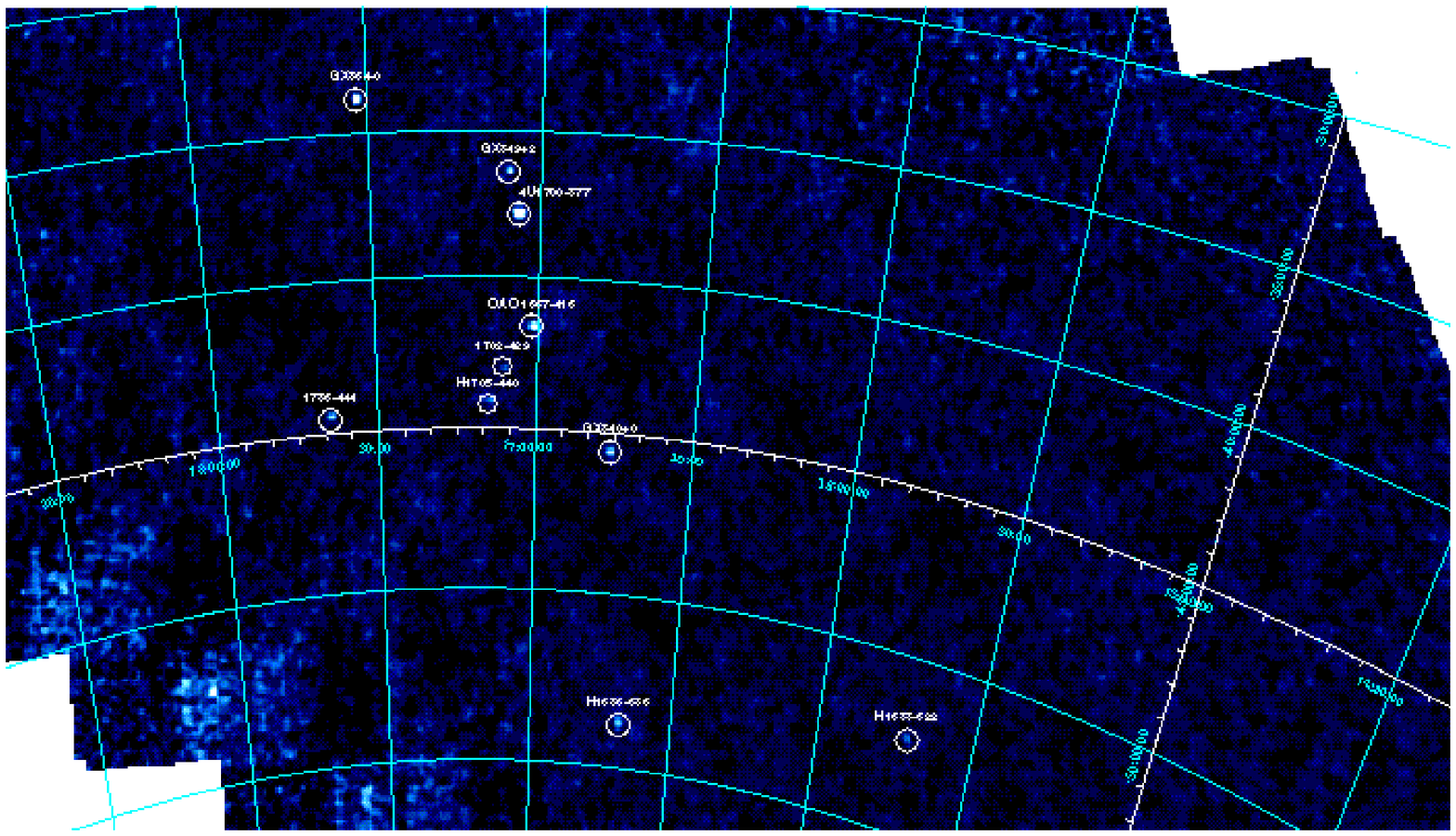}
      \caption{Skymap (equatorial J2000 coordinates) obtained by IBIS/ISGRI (15--30) keV, during Galactic plane scan on 12 March 2003. 
              }
         \label{ibismap}
   \end{figure*}

As an example of a GPS scan obtained by IBIS, Fig~\ref{ibismap} was obtained by IBIS/ISGRI during revolution 50 (12 March 2003), energy range (15--30) keV.
%
%
%

\section{First results from JEM-X }
During the GPS scans executed during the first half of 2003
JEM$-$X detected more than 50 known sources, most of them galactic
X$-$ray binaries. We also recorded numerous X$-$ray bursts. Most of
the sources had fluxes in excess of 20 mCrab. However, sources
down to 3 mCrab have been detected.

\begin{figure*}[htp]
   \centering
   \includegraphics[width=\textwidth]{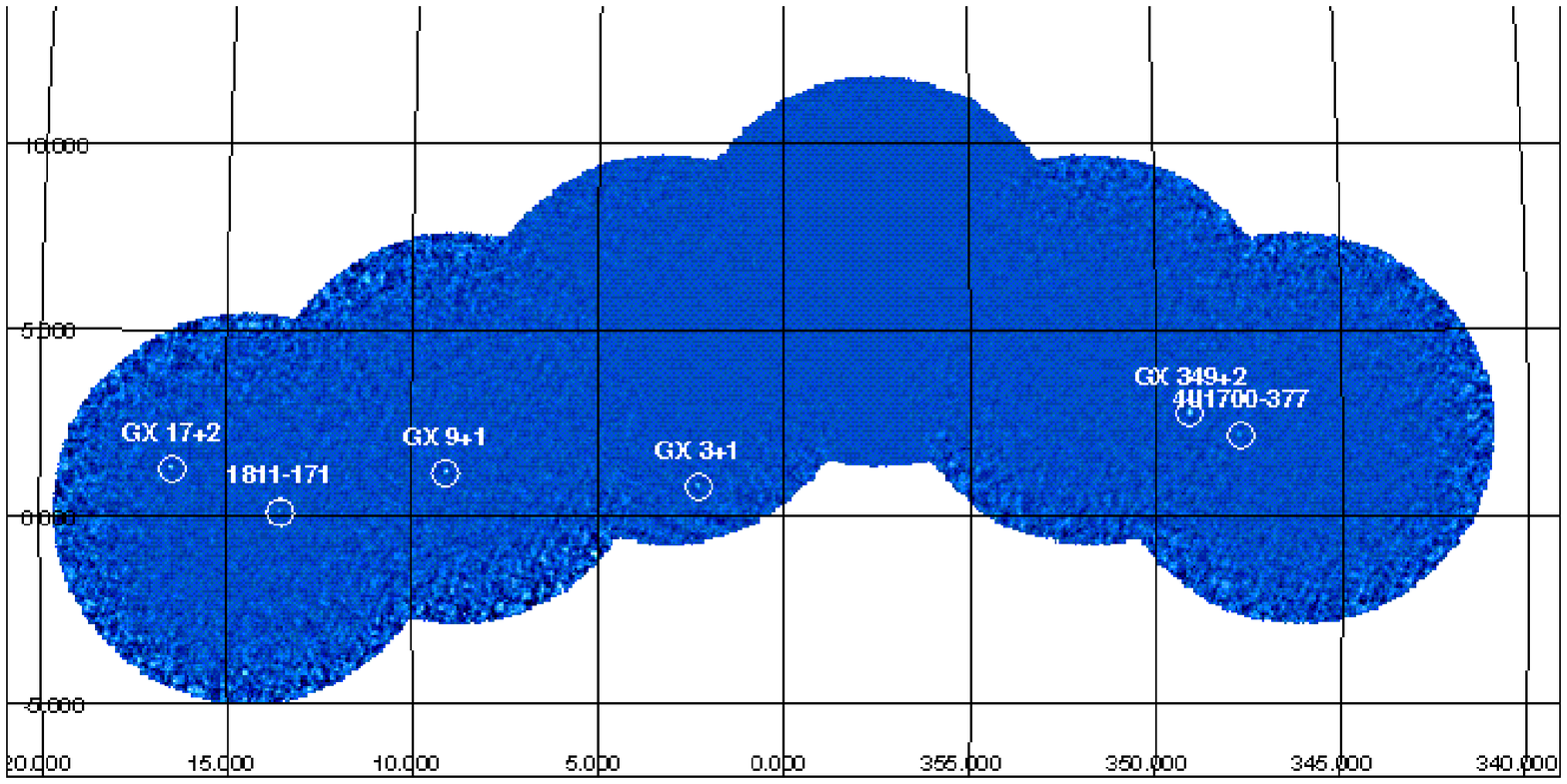}
      \caption{JEM$-$X sky map (3$-$15) keV of a GPS (24 March 2003) close to the Galactic Centre, see text. 
              }
         \label{jemxmap}
   \end{figure*}

Fig~\ref{jemxmap} shows a (3$-$15) keV sky map close to the Galactic Centre, 
obtained during the GPS on 24 March 2003. 
Six pointings of 2200 s each have been combined into a mosaic using the overall exposure and a map of the vignetting function to determine weights of summed pixels. 

\begin{figure*}[htp]
   \centering
   \includegraphics[width=14cm]{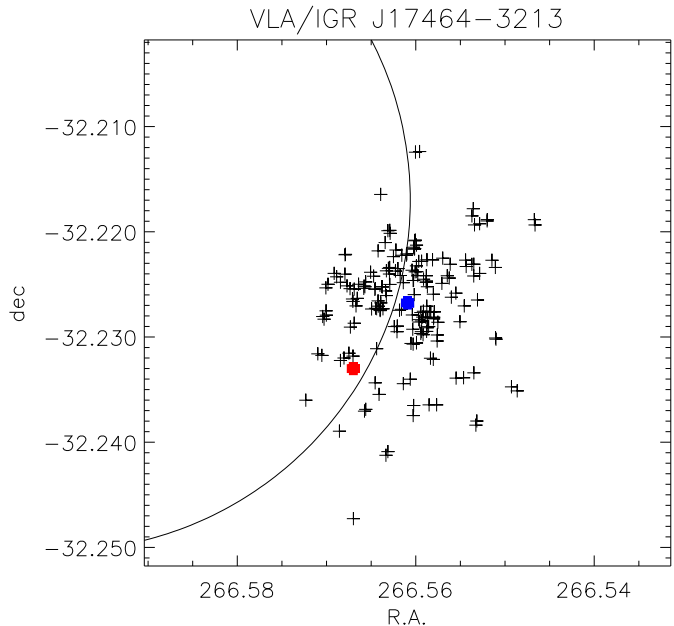}
      \caption{Original IBIS location error circle, counterpart suggested by
VLA observations (red dot) and JEM$-$X positions obtained from
many pointings and energy bands (black crosses). The JEM$-$X data
confirm the association of the radio and the hard X$-$ray source.
The blue dot marks the centroid of the JEM$-$X positions. A residual 
uncorrected error in the JEM$-$X alignment is evident.              }
         \label{VLA}
   \end{figure*}

So far, JEM$-$X has not detected a bona fide transient source which
was confirmed by independent observations. This  may not be
surprising, given the more restricted (partially coded) 
field of view (10$^{\circ}$ diameter) compared to the IBIS partially coded FOV 
of  29$^{\circ}$$\times$29$^{\circ}$. One of the INTEGRAL 
transients, IGR J17464$-$3213, was, however, detected a number of 
times after the initial discovery by IBIS. For this source, a HEAO source, H1743--322, which had its last outburst in 1977, detected by IBIS on 
2003, Mar 21 (Revnivtsev et al. \cite{revnivtsev2}), JEM$-$X
could confirm a VLA$-$radio counterpart (Rupen et al. \cite{rupen})
near the edge of the original IBIS error box as the true origin of the X-ray signal (Fig~\ref{VLA}).
%
%

The JEM$-$X source detection software is still developing, and we
expect eventually to reach a detection threshold around 5 mCrab 
in a single GPS pointing of 2200 s. Given that the angular
separation between the GPS pointings is comparable to the JEM$-$X FOV,
we will frequently  get the opportunity to observe a given
source in only one GPS pointing. To achieve a reliable identification 
of transient sources it will therefore be very beneficial
to get both JEM$-$X units operating in parallel -- at the present 
time only one JEM$-$X is active, pending the conclusion of the
investigation of a long term drift of the gas gain of JEM$-$X2
discussed elsewhere in this volume (Brandt et al. \cite{brandt}).

 %
%
%

\section{Outlook}

    In this section the sensitivities achieved to date during GPS observations are presented  in Table~\ref{sensitivities} and used to
predict the scientific return expected from future scans.
For sensitivities, we consider individual GPS scans with exposure per point of 2200 s.  These are the sensitivities relevant for transient detection and analysis.  The sensitivities are based on
actual source detections achieved since launch.  The IBIS number is for the low$-$energy ISGRI portion of the
response and relevant for single pointings (2200 s) in a scan, including systematic errors.  JEM$-$X is also quoted for single pointings.  For SPI, with its wide field of view, the sensitivity is derived from analysis of sources seen in a single GPS scan: a source is seen $\sim$3 times in a scan, hence the average exposure is $\sim$6600 s, depending on the source position with respect to the scan path.

   \begin{table}[htp]
      \caption[]{Sensitivities achieved during GPS observations (see text).}
         \label{sensitivities}
     
        \begin{tabular}{llll}

            \hline
            \noalign{\smallskip}
           Instrument &  $\Delta$E & 5$\sigma$ Sensitivity & Exposure \\
	    & [keV]&    [mCrab] & [ksec]\\
            \noalign{\smallskip}
            \hline
            \noalign{\smallskip}
           
   	IBIS &           15 $-$ 40 &         36 & 2.2 \\
   	SPI  &        20 $-$ 40   &       62  & 6.6 \\
   	JEM$-$X &          5 $-$ 20   &       20 & 2.2 \\

   \noalign{\smallskip}
            \hline
         \end{tabular}
    \end{table}

 For transient science, studies before launch gave a prediction of $\sim$10 transients detected per year by
INTEGRAL.  This was based on an estimated sensitivity threshold of $\sim$20 mCrab.  The classes of object
considered were:
(i) X$-$ray novae (including systems with black hole and with neutron star primaries), (ii) Be pulsar outbursts, (iii) AGN flares, and (iv) Novae and Supernovae.

    One of the great successes of the INTEGRAL mission to date has indeed been its discovery of new
transient sources (see Table~\ref{new}).  Ten new sources have been found over a period of about 4 months,
giving a discovery rate of more than 2 per month.  Most of those sources have been found during core programme observations, mostly during GCDE observations which have dominated the core programme time so far.  Combining the GCDE exposures which are clustered twice per year because of the visibility of the Galactic Centre region, and the once$-$per$-$12$-$day GPS scans, it is expected
that a rate of 10$-$20 transient discoveries per year will be obtained.  The range of uncertainty is due to not
knowing which of the new sources are steady and which are transient.

    For mapping science with SPI, the time spent on GPS scans to date (i.e. until May 2003: 282 pointings with a total of 643 ks observing time) has not yet been long enough to build sufficient statistics for diffuse continuum and line studies.  However, when
this number grows to $\sim$100 ks per GPS pointing at the end of the first year for SPI, it will be possible to start making maps of
the full plane.  The results from this should give sensitivities comparable to those achieved in prime locations
of the plane by the COMPTEL instrument on CGRO.  If such scans are continued over a 5 year extended
mission for INTEGRAL, the exposure per point will become 0.5 Ms.  The result will be a beautiful map of the
plane showing the sites of nucleosynthesis in the Galaxy over the past million years.

\section{Conclusions}
The Galactic plane scans (GPS) of INTEGRAL's core observing programme, executed so far during the early mission phase, have proven to be of fundamental importance for the science return of the mission. INTEGRAL is demonstrating its unique capability to routinely monitor the Galactic plane with large field of view and excellent sensitivity. The number of new sources found so far is very encouraging. The prospects of detecting transient line features and producing maps in the light of diffuse continuum and line emission are very promising.
The present structure of the core programme of INTEGRAL will in principle continue during the future operational lifetime.

\begin{acknowledgements}
      
The authors gratefully acknowledge the INTEGRAL Science Working Team responsible for the definition of the core observing programme, and the staff of the INTEGRAL Science Operations and Science Data Centres for the scheduling, execution, and data analysis support.

\end{acknowledgements}

\end{document}